\documentclass[sigconf]{acmart}

\usepackage{blindtext}
\usepackage{amsmath}
\usepackage{amssymb}
\usepackage{amsthm}
\usepackage[utf8]{inputenc}
\usepackage{todonotes}
\usepackage{graphicx}
\usepackage{subfig}
\usepackage{footnote}
\usepackage{booktabs}
\usepackage[export]{adjustbox}
\usepackage{xfrac}
\usepackage[keeplastbox]{flushend}
\usepackage{listings}
\usepackage{xcolor}
\usepackage{soul}
\usepackage{url}

\definecolor{mygreen}{rgb}{0,0.6,0}
\definecolor{mygray}{rgb}{0.5,0.5,0.5}
\definecolor{mymauve}{rgb}{0.58,0,0.82}
\definecolor{editorgray}{rgb}{0.95, 0.95, 0.95}

\sethlcolor{editorgray}
\let\OldTexttt\texttt
\renewcommand{\texttt}[1]{\OldTexttt{\hl{#1}}}

\renewcommand\footnotetextcopyrightpermission[1]{} 

\begin{document}
	
	\title{\vspace*{-1.5cm}fiction: An Open Source Framework for\\the Design of Field-coupled Nanocomputing Circuits\\\normalfont{(Extended Abstract)}}

	\author{%
		Marcel Walter$^1$ \quad Robert Wille$^{2,3}$ \quad Frank Sill Torres$^{1,3}$ \quad Daniel Große$^{1,3}$ \quad Rolf Drechsler$^{1,3}$\\[1ex]%
		$^1$Group of Computer Architecture, University of Bremen, Germany\\%
		$^2$Johannes Kepler University Linz, Austria\\%
		$^3$Cyber Physical Systems, DFKI GmbH, Bremen, Germany\\%
		\{m\_walter, frasillt, grosse, drechsler\}@uni-bremen.de, robert.wille@jku.at\\[1ex]%
        \emph{\url{https://github.com/marcelwa/fiction}}%
	}
	
	
	\pagestyle{plain}
	\pagenumbering{gobble}
	
	\begin{abstract}
		As a class of emerging post-CMOS technologies, \emph{Field-coupled Nano\-computing} (FCN) devices promise computation with tremendously low energy dissipation. Even though ground breaking advances in several physical implementations like \emph{Quantum-dot Cellular Automata} (QCA) or \emph{Nanomagnet Logic} (NML) have been made in the last couple of years, design automation for FCN is still in its infancy and often still relies on manual labor. In this paper, we present an open source framework called \emph{fiction} for physical design and technology mapping of FCN circuits. Its efficient data structures, state-of-the-art algorithms, and extensibility provide a basis for future research in the community.
	\end{abstract}

	\thanks{This paper discusses \emph{fiction} v0.2.1}

	\maketitle
	
	\section{Introduction and Background}\label{sec:intro}
	
	\emph{Field-coupled Nanocomputing}~(FCN)~\cite{Anderson14} is a class of emerging technologies which conduct computations fundamentally differently from conventional systems relying e.\,g.~on CMOS. Here, information is represented in terms of the polarity or magnetization of nanoscale cells and can be propagated to adjacent ones using repelling forces of local fields~\cite{Lent97, Giri18}. This results in devices that allow to represent and process binary information without electrical current flow. Consequently, numerous contributions on their physical realization have been made in the past and several of some them in the last three to four years, e.\,g. \emph{molecular Quantum-dot Cellular Automata}~(mQCA)~\cite{Lent16}, \emph{atomic Quantum-dot Cellular Automata}~(aQCA)~\cite{Bohloul17, Huff17}, or \emph{Nanomagnet Logic}~(NML)~\cite{Hu15}.
	
	Moreover, this way of representing and processing information is doable with highest processing performance and remarkably low energy dissipation -- as confirmed by several theoretical and experimental studies (see e.\,g.~\cite{Timler02, Pitters11,Sill18}).
	This makes FCN a promising alternative to conventional integrated circuit technologies. However, no exhaustive automatic design flow is available for FCN technologies so far. Also, due to different design rules of CMOS VLSI, existing classical methods are not applicable to the FCN domain.
	
	In this paper, we present \emph{fiction}, a framework for the design of FCN circuits. The framework is written in C++ and uses the \emph{EPFL Logic Synthesis Libraries}~\cite{EPFLLibraries}. With this tool, we especially tackle the physical design steps of FCN like placement, routing, timing, and technology mapping under the domain specific constraints.
	
	Logic synthesis however is taken as granted as it can be performed with existing tools like \emph{ABC}~\cite{brayton2010abc}. Even though their physical implementations differ from each other, the structural models of most FCN technologies are nearly identical. Data structures in \emph{fiction} are designed around this insight: whenever possible, \emph{fiction} abstracts from physical implementations and conducts layout on a higher level. Only in the final step, a technology mapping is performed.
	
\begin{figure}[!tp]
		\centering
		\subfloat[AND\label{fig:and}]{\includegraphics[scale=.5]{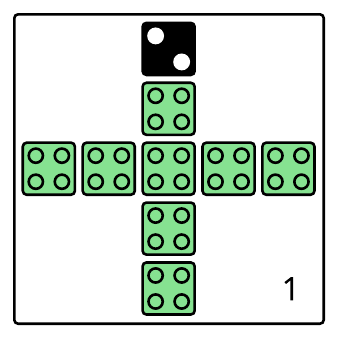}} \quad
		\subfloat[INV\label{fig:inv}]{\includegraphics[scale=.5]{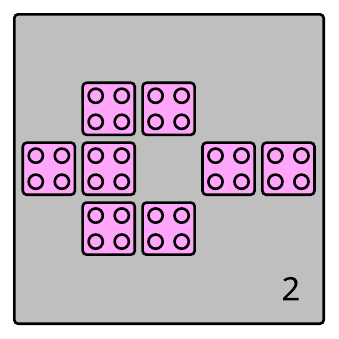}} \quad
		\subfloat[Wire\label{fig:wire}]{\includegraphics[scale=.5]{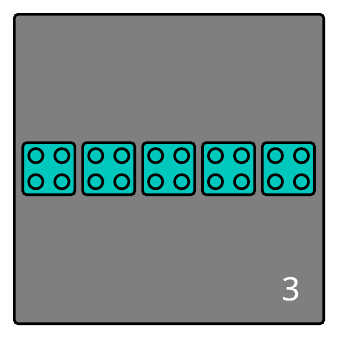}} \quad
		\subfloat[Fan-out\label{fig:fanout}]{\includegraphics[scale=.5]{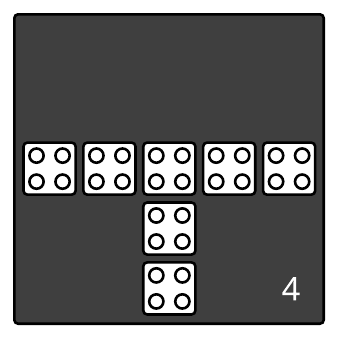}} 
		\caption{Tiles in QCA implementation}
		\label{fig:tiles}
	\end{figure}

	On that layer of abstraction, the design task boils down to the composition of tiles with assigned logic or wire elements. Such entities for an AND gate, an inverter, a straight wire, and a fan-out are shown exemplarily in Figure~\ref{fig:tiles}. The shade of the tiles, the coloring of the logic elements, and the numbers in the bottom right corners represent redundant information about the clocking. For further information see~\cite{Hennessy01, Walter2018, Anderson14, Sill18}.
	
	In the following, \emph{fiction} is initially presented from the perspective of a standard user in Section~\ref{sec:user}, where an example layout flow is conducted and benchmarking is elaborated, followed by a description of the developer's perspective in Section~\ref{sec:dev}, where the implementation of a naive random placement is exemplarily shown. Section~\ref{sec:concl} concludes the paper.

	\section{The User's Perspective}\label{sec:user}
	
	In this section, two typical application scenarios within \emph{fiction} are described. First, it is shown how interaction with the store-based command-line interface (CLI) \emph{alice}~\cite{EPFLLibraries} works by the use case of obtaining a routed layout which is prepared for physical simulation. Then, scripting, benchmarking and logging functionalities to easily generate statistical data are demonstrated.
	
	\subsection{The CLI}\label{sec:cli}
	
	Starting point of all flows is a synthesized \emph{Verilog} netlist file which exclusively uses the assign statement and logic primitives. Suitable files can be generated with \emph{ABC}~\cite{brayton2010abc} using the following commands.
	
\begin{lstlisting}
read <inputfile>
strash
write <outputfile>.v
\end{lstlisting}
	
	Also, \emph{fiction} comes with a set of sample netlist files which can be found in the \mbox{\texttt{benchmarks}} folder. These can be loaded using the command~\texttt{read}. By this, the netlist is parsed and placed in a \emph{store} where it can be accessed and (re-)used by other algorithms. So far, two state-of-the-art layout approaches are implemented which can be called via the commands~\texttt{exact}~\cite{Walter2018} and \texttt{ortho}~\cite{walter2019scalable}. Algorithm \texttt{exact} utilizes the \textsc{Smt} solver~\emph{Z3}~\cite{DeMoura:2008} to generate minimal layouts in terms of area within the provided parameters. The approach is highly configurable and allows to toggle and set up several design criteria like the clocking scheme, the use of crossings, balanced paths, synchronization elements~\cite{torres2018synchronization}, I/O pins, and wire length restrictions. However, \texttt{exact} is only applicable for rather small netlists due to the complexity of the tackled problem~\cite{walter2019np}. On the other hand, \texttt{ortho} is a heuristic algorithm which does not guarantee minimal layouts anymore and also is restricted to a fixed clocking scheme, but therefore can generate results in feasible runtime.
	
	Eventually, both algorithms generate a gate-level abstraction of an FCN circuit grid. Using the command~\texttt{cell}, a technology mapping is performed with a selected gate library, whose default is \emph{QCA-ONE}~\cite{Reis16}. Having a QCA circuit in store, it can be written as a simulation file for the \emph{QCADesigner}~\cite{walus2004qcadesigner}, a standard tool for physical simulation of QCA structures, by entering~\mbox{\texttt{qca <filename>.qca}}. Also, using the command~\texttt{show} generates a scalable vector graphic to inspect the implemented circuit.
	
	\subsection{Benchmarking \& Scripting}\label{sec:bench}
	
	The flow shown in the previous section can easily be repeated by storing it in a \emph{fiction script} file. We assume that two designs generated with different settings of the \texttt{exact} algorithm for the netlist \texttt{c17.v} shall be compared. Therefore, we create the following file \texttt{compare.fs}.
	
\begin{lstlisting}
read ../benchmarks/ISCAS85/c17.v
exact -ixbs 2ddwave4
ps -g
cell
show
exact -ps use
ps -g
cell
show
\end{lstlisting}
	
	The first call to \texttt{exact} enables designated I/O pins~(\texttt{-i}), allows crossings~(\texttt{-x}), routes all I/Os to the grid borders~(\texttt{-b}) and uses the 4-phase \emph{2DDWave}~\cite{vankamamidi2006clocking} scheme~(\texttt{-s 2ddwave4}), while the second one allows for unbalanced (de-synchronized) paths~(\texttt{-p}) and utilizes \emph{USE}~\cite{Campos16} as the clocking scheme~(\texttt{-s use}).\footnote{Further predefined clocking schemes include \emph{RES}~\cite{goswami2019efficient} and \emph{BANCS}~\cite{formigoni2018bancs}. Though the default is an irregular open clocking which gives the solver a degree of freedom in assigning the clock numbers itself.}

\begin{figure}[!tp]
		\centering
		\subfloat[exact -ixbs 2ddwave4\label{fig:compare1}]{\includegraphics[scale=.25]{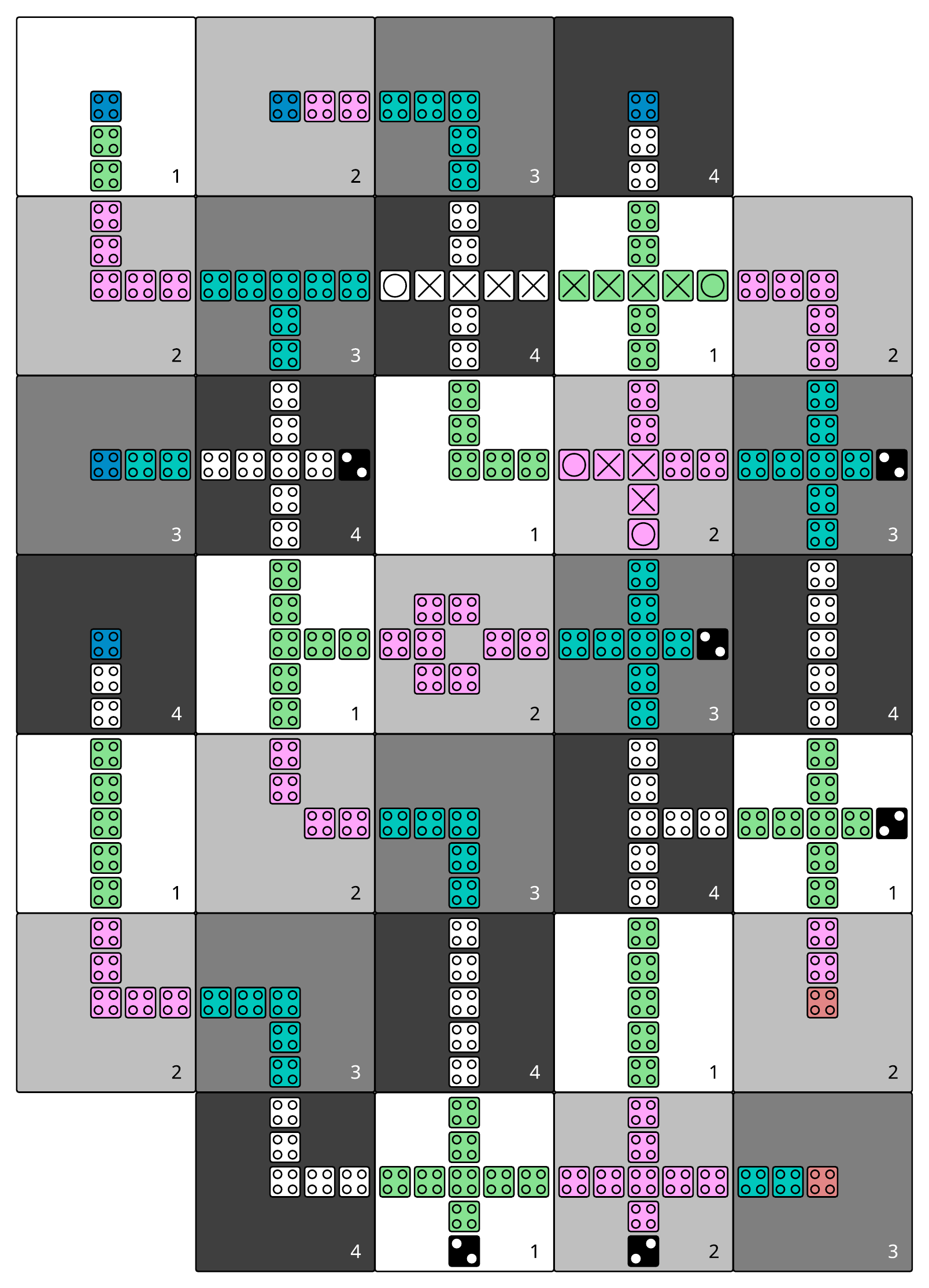}} \qquad
		\subfloat[exact -ps use\label{fig:compare2}]{\includegraphics[scale=.25]{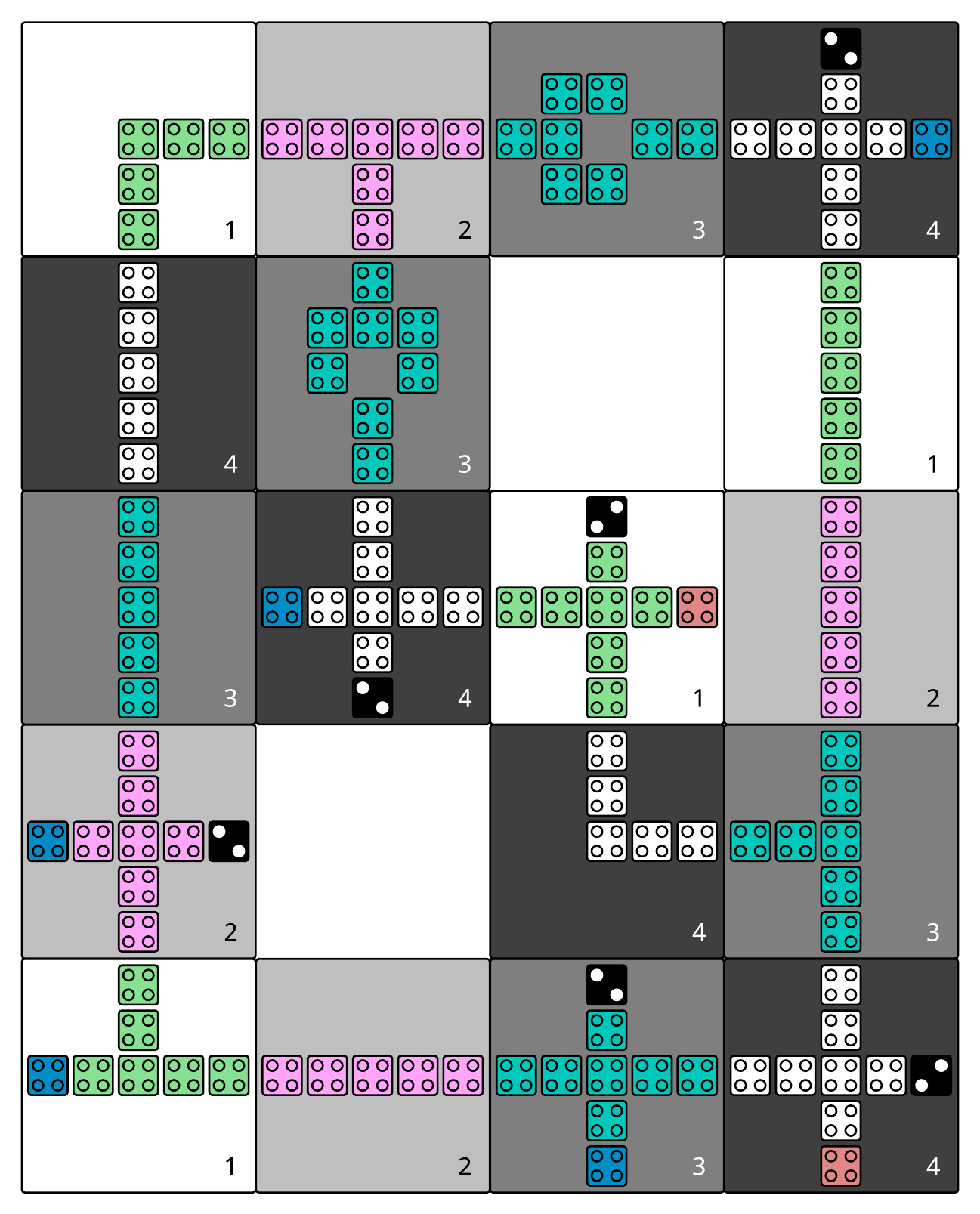}}
		\caption{Two differently layouted variants of c17.v}
		\label{fig:exact}
	\end{figure}

	To run this script, we enter \texttt{./fiction -f compare.fs}. Not only do we get SVG images of both layouts but also, through the use of \texttt{ps~-g}, some statistical information about the layouts are printed, i.\,e. the dimension of the resulting grid in tiles, the amount of gate~(\texttt{\#G}) and wire tiles~(\texttt{\#W}), crossings~(\texttt{\#C}), and latches~(\texttt{\#L}) used, the length of the critical path~(\texttt{CP}) in tiles, and the throughput~(\texttt{TP})
	~\cite{torres2018synchronization, Sill18DSD}. Note that fan-outs and I/O pins are counted as gates since they are fixed by the input. This way, the displayed amount of wires represents the net costs~\cite{Walter2018dsd}. The resulting graphics are shown in Figure~\ref{fig:exact}.
	
\begin{lstlisting}
c17: 5 x 7, #G: 18, #W: 18, #C: 3, #L: 0, CP: 11, TP: 1/1
c17: 4 x 5, #G: 11, #W:  7, #C: 0, #L: 0, CP: 13, TP: 1/3
\end{lstlisting}

	Finally, these functionalities can be embedded into a shell script. For the next scenario, we want to layout all files from a folder, log their statistical information, and generate simulation models for QCADesigner. To this end, we create the following bash script.
	
\begin{lstlisting}[language=bash]
for filepath in ../benchmarks/TOY/*.v; do
  f="${filepath##*/}"
  ./fiction -c "read $filepath; ortho; ps -g; cell; qca  ${f%.*}.qca" -l ${f%.*}.json
done
\end{lstlisting}

	Using the \texttt{-c} flag, a semicolon-separated list of commands can be provided and the output is logged in a JSON file by the \texttt{-l} flag. Note that when logging is activated, \texttt{ps -g} logs contain more in-depth data about the layout like its bounding box size, energy consumption~\cite{Sill18}, etc. For both, the physical models as well as the log files, the originally entered file name is used extended by the respective file extension.
	
	\section{The Developer's Perspective}\label{sec:dev}
	
	This section elaborates important design decisions for \emph{fiction}'s data types and presents some sample code for a naive random placement in order to demonstrate their use.
	
	Core of the implementation are the classes \texttt{fcn\_gate\_layout} and \texttt{fcn\_cell\_layout} -- a gate level, tile-based abstraction and a physical FCN cell-based layout respectively. They are based on a \texttt{boost::grid\_graph}, a highly memory efficient grid data structure from the \emph{Boost Graph Library}~(BGL)~\cite{siek2002boost}. Following BGL's paradigm, the grid topology is separated from the associated elements like wires, gates, or cells respectively. Such associations happen via defaulted maps that return standard values (mostly used for free grid positions), when an uninitialized access happen to save even more memory.
	
	Also, regular clocking schemes like USE are stored in terms of a small cut-out which is then seamlessly extrapolated for larger layouts as there is no need to store clock values for every single tile. Furthermore, all data structures provide convenience functions and iterators to shift attention away from implementation details and towards actual algorithms when working with \emph{fiction} as a developer. 
	
	The following code snippet demonstrates how a simple function can be implemented to randomly place vertices of a \texttt{logic\_network} (constructed from a parsed Verilog file) on layout tiles.\footnote{Note that applicable usages of \texttt{std::move} have been omitted due to space limitations.\label{foot:move}}
	
\begin{lstlisting}[language=c++]
void naive_random_placement()
{
  // fetch current logic network from store
  auto network = store<logic_network_ptr>().current();
  auto n = network->vertex_count();
  
  // create an empty 4-phase USE layout of size n x n
  auto layout = std::make_shared<fcn_gate_layout>(fcn_dimension_xy{n, n}, use_4_clocking, network);

  // for all logic vertices v
  for (auto&& v : network->vertices())
  {
    auto placed_successfully = false;
    do
    {
      // sample a random tile t in ground layer
      auto t = layout->random_tile(GROUND);
      if (layout->is_free_tile(t))
      {
        // place v at t
        layout->assign_logic_vertex(t, v);
        placed_successfully = true;
      }
    } while (!placed_successfully);
  }
  // place resulting layout in a store
  store<fcn_gate_layout_ptr>().extend() = layout;
}
\end{lstlisting}
	
	The given function \texttt{naive\_random\_placement} can for instance be implemented as a new command in the file \texttt{io/commands.h} by following the scheme of existing ones or by considering the official \emph{alice} documentation. Note that the function as given here only places gates but does neither take care of their orientation nor their routing. A custom router can fully benefit from the whole functionality offered by the BGL as most of their (path finding) algorithms work on \texttt{boost::grid\_graph}s as well.
	
	Assuming the routing step has happened as well, one might want to convert the gate level abstraction to an actual cell-based implementation to conduct lower level optimizations. We further assume, the QCA-ONE gate library should be utilized for the technology mapping. The following code snippet does the job.\textsuperscript{\ref{foot:move}}
	
\begin{lstlisting}[language=c++]
void technology_mapping()
{
  // fetch current gate layout from store
  auto gates = store<fcn_gate_layout_ptr>().current();
  // prepare a library object for technology mapping
  auto lib = std::make_shared<qca_one_library>(gates);
  // apply library to generate a cell-based layout
  auto cells = std::make_shared<fcn_cell_layout>(lib);
  // store cell layout
  store<fcn_cell_layout_ptr>().extend() = cells;
}
\end{lstlisting}

	For further information, we refer the reader to \texttt{io/commands.h} and \texttt{tech/fcn\_gate\_library.h}.
	
	Additionally to the already introduced functionalities, \emph{fiction} supports the use of externally clocked synchronization elements~\cite{torres2018synchronization}, the use of multiple wires elements in the same tile in the gate-level abstraction already, and the direct construction of (cell-wise clocked) physical cell layouts without the use of any gate library.	
	
	\section{Conclusion}\label{sec:concl}
	
	In this paper, we introduced \emph{fiction}, an extensible open source framework written in C++ for the layout, optimization, and physical design of Field-coupled Nanocomputing Circuits. The framework comes with efficient data structures, state-of-the-art algorithms, as well as rich scripting and logging functionalities. We thereby provide a foundation for future research in the community.
	
	\begin{acks}
		We would like to thank Gregor Kuhn and Mario Kneidinger for code contributions and the authors of the \emph{EPFL Logic Synthesis Libraries} for their inspiring work.
	\end{acks}


	\makeatletter
	\newcommand*\mysize{%
	  \@setfontsize\mysize{6.4}{6.6}%
	}
	\makeatother
%
	%

    \renewcommand*{\bibfont}{\mysize}
	\bibliographystyle{ACM-Reference-Format}
	\bibliography{lit_header_short,bibliography}

\end{document}